\documentclass[useAMS,usegraphicx]{mn2e}

\def\ltsima{$\; \buildrel < \over \sim \;$}
\def\lsim{\lower.5ex\hbox{\ltsima}}
\def\gtsima{$\; \buildrel > \over \sim \;$}
\def\gsim{\lower.5ex\hbox{\gtsima}}

\newcommand{\be}{\begin{equation}}
\newcommand{\en}{\end{equation}}

\begin{document}
   \title{A wide field X--ray telescope for astronomical survey purposes: from theory to practice}

\author[P. Conconi et al.]{Paolo Conconi$^{1}$, Sergio Campana$^{1,}$\thanks{E-mail: sergio.campana@brera.inaf.it}, 
Gianpiero Tagliaferri$^{1}$,  Giovanni Pareschi$^{1}$,
\newauthor Oberto Citterio$^{1}$,
Vincenzo Cotroneo$^{1}$, Laura Proserpio$^{1}$, Marta Civitani$^1$\\
$^1$ INAF-Osservatorio Astronomico di Brera, Via Bianchi 46, I--23807, Merate (LC), Italy}

\maketitle

\begin{abstract}
X--ray mirrors are usually built in the Wolter I (paraboloid-hyperboloid) configuration. 
This design exhibits no spherical aberration on-axis but suffers from field curvature, coma and 
astigmatism, therefore the angular resolution degrades rapidly with increasing off-axis angles. 
Different mirror designs exist in which the primary and secondary mirror profiles are expanded 
as a power series in order to increase the angular resolution at large off-axis positions, at the expanses
of the on-axis performances. Here we present the design and global trade off study of an X--ray mirror 
systems based on polynomial optics in view of the Wide Field X-ray Telescope (WFXT) mission. 
WFXT aims at performing an extended cosmological survey in the soft X--ray band with unprecedented 
flux sensitivity. To achieve these goals the angular resolution required for the mission is very demanding 
$\sim 5$ arcsec mean resolution across a 1-deg field of view. In addition an effective area 
of 5--9000 cm$^2$ at 1 keV is needed. 
\end{abstract}


\section{Introduction}

Focussing telescopes for X--ray astronomy are usually built in the Wolter I configuration (Wolter 1952a, 1952b) 
providing, at least theoretically, perfect images for on-axis sources. Wolter I telescopes are made by
two mirror segments, shaped as two surfaces of revolution (with a parabolic and hyperbolic profile, respectively),
joining at the intersection plane. 
Despite theoretically perfect images on-axis, the image quality rapidly degrades far from the optical axis due to the curvature
of the best focal plane, spherical and chromatic aberrations, 
limiting the capabilities of carrying out surveys of the X--ray sky. 
Simple solutions were suggested to improve the off-axis angular response, like shifting slightly out of focus a single detector 
(e.g. Cash et al. 1979) to get closer to the best focal plane at larger off-axis angles (like with the SWIFT XRT 
CCD, Burrows et al. 2005) or tilting the detectors (if more than one) to better approximate the (curved) best focal plane 
(as with the Chandra ACIS-I detectors or the XMM-Newton MOS detectors).
These simple recipes however provide only mild improvements. In order to sensibly improve the off-axis response
of an X--ray telescope  it is mandatory to act directly on the mirror design. 
The Wolter-Schwarzschild telescope eliminates the coma aberration for paraxial rays and provides a superior 
response at large off-axis angles with respect to the Wolter I telescope (Chase \& VanSpeybroeck 1973).
Nariai (1987, 1988) suggested the idea of a telescope made by two hyperboloid surfaces (see also Harvey \&
Thompson 1999) which provides good performances over a field of view of $\sim 20'$. This was adopted for 
the Solar X--ray Imager telescope.
Mirror shells for X--ray telescopes can be built, with the 
same degree of complexity, with polynomial profiles (Werner 1977; Burrows, Burg \& Giacconi 1992, hereafter BBG).
Polynomial mirror profiles are described usually by forth (or third) grade polynomia and 
optimization techniques can be easily implemented (BBG; Conconi \& Campana 2001, CC hereafter).
Depending on the a priori optimization specifications one can re-discover the Wolter I design (optimizing for the 
best on-axis angular response) or other designs optimizing the angular response over a given field of view.
Wide field design are commonly used for X--ray observations of the Sun (e.g. Tsuneta et al. 2000; Lemen et al. 2004;
DeLuca et al. 2005) and have been proposed for X--ray survey purposes  (e.g. BBG, CC; Conconi et al. 2004; 
Thompson \& Harvey 2000).

Besides changing the mirror shape, additional improvements can be introduced. In particular, as also discussed in CC, 
an improvement of the optical quality can be achieved by building different shells with a variable mirror length dependent 
upon the radius, giving to the total mirror assembly a butterfly-like shape. 
This solution causes a reduction of the total effective area with respect to shells with equal (longer) length, but it allows 
to keep the same curvature of the focal plane for the different shells, with the consequent improvement of image quality
and sensitivity. 
In addition, mirror shells have a different plate scale, so that at any off-axis angle the X--rays of each shell are focussed 
in slightly different positions. To correct for this effect shells have to be built with modified focal lengths and (small) displacements at the intersection planes must be introduced to compensate for this.
Modified focal lengths and small displacements of the intersection planes were introduced on Chandra to keep 
a uniform plate scale.
With these simple recipes any survey exposure time can be reduced by a factor of $\sim 2$ (CC).

In this paper we improve the optimization scheme performed by CC, making it more general.
In Section 2 the general theory to optimize one single mirror shell for 
a wide field of view is described. 
Section 3 deals with the optimization of a full mirror module, composed by several mirror shells. 
In order to reduce the number of free parameters, some assumptions on the system geometry are considered.
Since the parametrization hereby adopted is based on dimensionless units, a fixed arbitrary focal length can be used 
for the simulations. For the application to a real telescope a scale must be set, determining the physical size of the system.
The imaging performances of the telescope are not affected by the scaling, that determines instead areas, volumes and masses. 
In Section 4 we consider the application of the developed concepts to the design of the telescope for the proposed
Wide Field X--ray Telescope mission. In Section 5 we will briefly discuss the scientific capabilities of this telescope 
for the realization of wide area X--ray surveys.

\begin{figure*}
\label{telesco}
\centerline{
\includegraphics[width=15cm]{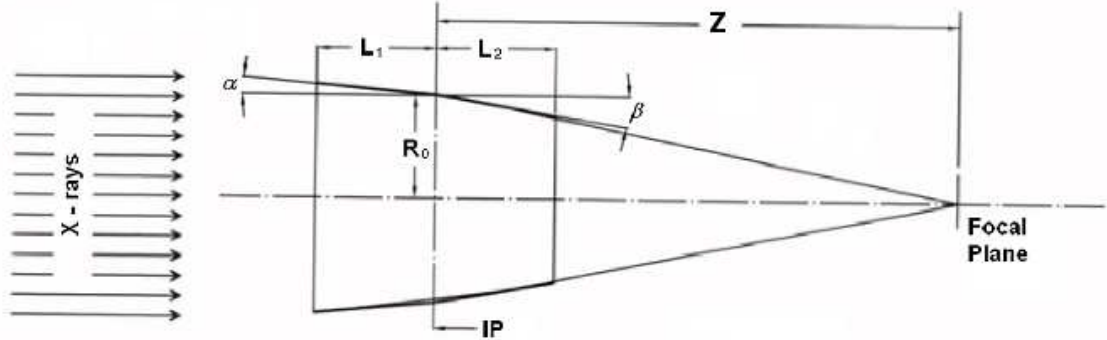}}
 \caption{Telescope configuration and parameters. $R_0$ is the mirror shell radius at the intersection plane (IP); 
$\alpha$ ($\beta$) is the angle the between the primary (secondary) mirror tangent at the intersection plane and 
the optical axis; $L_1$ ($L_2$) is the length of the parabola (hyperbola) mirror segment; $Z$ is the distance between 
the IP and the focal plane along the optical axis. For an X--ray telescope $Z$ is very close to the focal length of 
the telescope $F$.
}
\end{figure*}

\section{Optimal design of a single mirror shell}

The optimal design of a single mirror shell depends on the scientific aim one wants to pursue. In particular, some boundary 
conditions to the problem can be specified from considerations about the field of view. In the present case we assume 
a circle with radius $30'$ (0.8 square degrees). In a real telescope, a shell is partially obscured by the next innermost one, 
dependent upon the spacing between the two. It is important to consider this parameter from the beginning, since it affects 
the results of the optimization by changing the effective area at any off-axis angle. 
The spacing between two adjacent shells is defined, in this work, in terms of their acceptance angle, 
which is the angle made from the radius at the intersection plane of the outer shell to the outer radius of the inner 
shell at the edge of the first surface (an acceptance angle of $0'$ 
means that the outer radius of the inner shell is equal to the radius of the outer shell at the intersection plane). 
Larger acceptance angles make possible to see the sky from the intersection plane.
We assume an acceptance angle of $17'$, that is shown in Section 3 to be the optimal value for the whole telescope in 
the soft X--ray band.
The optimization is performed by searching for the mirror coefficients, that are able to minimize a suitable merit function $M$,
representative of the imaging quality over the field of view (the lower the value, the better the imaging quality). 
For a sensitive X--ray survey it is important to have small source spots and, at the same time, a Point Spread Function 
(PSF) with small wings, in order to avoid spreading photons in the field of view, leading to an increase of the background, 
and the superposition of sources. Therefore we have built the merit function as  
the average of the $50\%$ (also known as Half Energy Width, HEW) and $80\%$ (Encircled Energy Fraction, EEF$_{80}$)
containment radii. Clearly there are no means to know if this is the best possible merit function, 
but the optimization of the core and of the wings PSF should be effective in detecting X--ray sources.

The resulting merit function for a single mirror shell is:
\begin{equation}
M_{\rm single}=\sum_{j=0}^{N}\,
({\rm HEW}(\Theta_j)+{\rm EEF}_{80}(\Theta_j))/2\times \Theta_j d\Theta_j \label{meritsingle}
\end{equation}

\noindent where the sum is evaluated over $N$ values of the off-axis angle  $\Theta_j$ (with step $d\Theta_j$). 
The image quality is evaluated on the best curved focal plane. 

Clearly we need also to specify some dimensions (see Fig. 1 for the geometry of the problem).
We choose to optimize a ``unitary" shell with focal length $F$ of one meter and
scale all the other geometrical parameters as dimensionless quantities. 
We define $f$ as the ratio of the focal length $F$ over the shell entrance diameter $D$   
and $l$ as 100 times the total mirror shell length $L_1+L_2$ (the lengths of the first and the second reflecting surfaces, 
$L_1$ and $L_2$, can be different, even if usually $L_1=L_2=L$, and the total length is $2\,L$) over the focal length $F$
(i.e. $l=100\times (L_1+L_2)/F$). 
The dimensional quantities (area, volumes, weights) are defined for a real mirror shell or telescope, by setting the scale 
of the system, i.e. by defining the focal length. 
The free parameters for the minimization of the merit function $M$ are the geometrical parameters $f$ and $l$, defining the 
size of the shell, and the 4+4 coefficients describing the polynomial profiles of the two mirror surfaces (see CC for more details).

Having in mind the building of a telescope we first explore the area to weight ratio for different mirror shells as a function of the focal 
ratio $f$. We found that the area at 1 keV to weight ratio has a broad plateau for $5\lsim f \lsim 7$. We also found that 
the area at 4 keV to weight ratio has a broad plateau for $12 \lsim f \lsim 16$. In addition,  for $f>17$ there are not advantages 
in using the polynomial profile with respect  to a Wolter profile. We then consider the interval $5-17$ for the focal ratio $f$ (for smaller
values of $f$ the contribution of the mirror shells to the total area at 4 keV would be minimal but it would be large in terms of weight).
We also put the working constraint of  $f\times l <100$.

In order to minimize $M$ we have implemented an IDL software code which uses a damped least squares (DLS) method 
(Hayford 1985). DLS is a downhill minimization method using derivatives to attain the best fit solution and it is 
widely used in many optical design programs.  
In order to overcome possible stagnation in local minima it is usually coupled with other optimization methods. 
We used a ``hammer'' technique to refine the best solution found by the DLS minimization performing a search over 
a $3\times3$ grid for each parameter iteratively, varying also the grid step, until convergence (variation 
of the merit function lower than a given threshold) is achieved. The value of the merit function for any given 
parameter search is evaluated by means of ray-tracing simulations. 

Following the optimization procedure we found that, for $10 \lsim f\times l$, the following  
approximate relation provides a good (difference smaller than $0.5''$) description of the mean HEW at 
1 keV weighted over the entire $30'$ radius field of view (and evaluated over the best curved focal plane individuated 
by the merit function above):
\begin{equation}
{\rm HEW} ({\rm 1 keV},\,f,\, l)\simeq 0.42 + 0.053 \times f \times l  \ {\rm arcsec} \label{hewsingle}
\end{equation}

\noindent It follows that the best single mirror shell will be the one with the lowest $f$ and/or $l$ and for the same value of $f\times l$ 
one gets the same mean angular resolution irrespective of any other parameter. An example of this can be found in Fig. 2. 
Clearly this is not fully realistic since errors are not specified. Fixing an error budget of, e.g., 3 arcsec on the mean HEW at 1 keV,
Eq. 2 tells us that any configuration with $f\times l \lsim 50$ will provide the same results.

This gives us the opportunity to further reduce the parameter space. Given $5<f<17$, being interested in optimal designs
we can also limit $l$ in order to have the mean HEW to be lower than $4''$ for  single mirror shell. 
Using Eq. 2 this provides $4\lsim l \lsim14$.
In practice, the manufacturing of short mirror shells is more difficult as the shell gets shorter, because radius errors 
of the longitudinal profiles increase (clearly the same slope error has a larger impact on a shorter shell). 

\begin{figure}
\label{fl50}
\centerline{
\includegraphics[width=6.0cm,angle=-90]{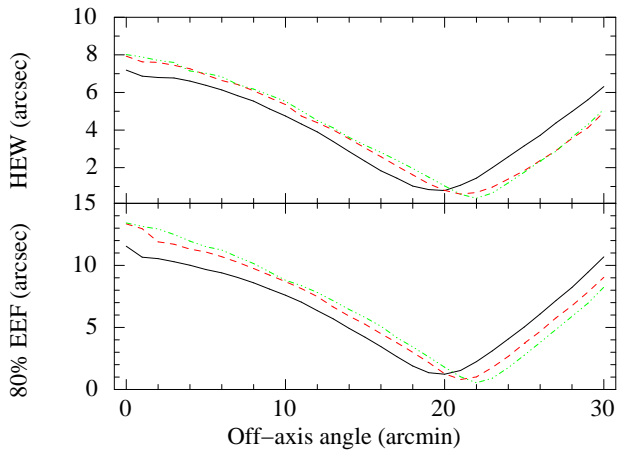}}
 \caption{Best image quality functions for a single mirror shell. The upper panel shows the HEW 
for three different mirror shells are reported as a function of the off-axis angle:
the continuous line refers to $f=5$ and $l=10$, the dashed line to $f=7.1$ and $l=7.1$, and the dash-dot-doted line to
$f=10$ and $l=5$. In the lower panel the $80\%$ EEF as a function of the off-axis angle for the same shells is shown.
Curves for the different mirror shell type have better performances on-axis for larger $l$ numbers. All the three
mirror shells have the same mean performances over the entire field of view.}
\end{figure}

Eq. 2 has been obtained evaluating the HEW at different angular distances from the optical axis 
on the best curved focal surface. According to the fit of our ray-tracing results, the radius of curvature of the focal plane 
$\delta$ can be expressed as:
\begin{equation}
\delta=10.9\times F /(f^{1.8}\times l) \label{curvasingle}
\end{equation}
Alternatively, the best curve focal plane can be described by the sagitta, i.e. the difference between the best focal 
length on-axis and the best focal length at any off-axis angle, can be expressed as:
\begin{equation}
S=0.046\times F \times l \times f^{1.8} \times \theta^2 \label{sagitta}
\end{equation}
\noindent (with $\theta$ the off-axis angle in arcmin).
This equation can be compared with the analog expression for the 
best curved focal plane for a Wolter I single shell (vanSpeybroeck \& Chase 1972), which 
can be rewritten as:
\begin{equation}
S_W=0.041\times F \times l \times f^{2} \times \theta^2  \label{curvasingle}
\end{equation}
indicating that the curvature of the focal plane for a polynomial mirror shell has a flatter 
dependence on the focal ratio $f$ than for a Wolter I shell. The last equation (i.e. Eq. 5) is equivalent to Eq. 6 in 
vanSpeybroeck \& Chase (1972), with the difference of having defined the best focal plane as the one with the best 
merit function (instead of the best r.m.s. spot) and for $5<f<17$ and $4<l<14$ (instead of $4<f<14$ and $7<l<35$). 
Simply rewriting their original equation results in a different value of 0.035 for the numerical coefficient.

One can also wonder if polynomial profiles are difficult to be manufactured, therefore providing a 
strong bias against their use in X--ray astronomy. In Fig. 3
we plot the differences with respect to a double cone mirror shell of the parabola and the hyperbola segments 
of a Wolter I design as well as the differences of the two segments of a polynomial mirror shell. 
This figure clearly shows that there is no manufacturing bias against polynomial mirror shells.
The idea of the polynomial designs in grazing incidence optics comes from optical astronomy 
with Ritchey-Chr\'etien Cassegrein telescope, were by deliberately compromising the on-axis performances,
one can introduce aberrations (mainly spherical) that tend to cancel or reduce the on-axis aberrations.
In Appendix A we carry out a detailed aberration analysis of our best fit mirror shell together with Wolter I and 
Wolter-Schwarzschild single mirror shells.

\begin{figure}
\label{para}
\centerline{
\includegraphics[width=6.0cm,angle=-90]{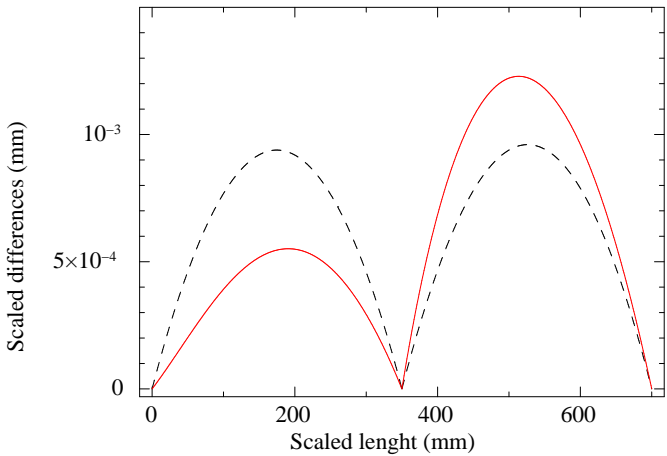}}
\caption{Differences between a Wolter I (dashed) or a polynomial (continuous) mirror shell with respect to a double cone 
profile for a $f=5$ and $l=7$ mirror and with focal length $F=1000$ mm. }
\end{figure}

\begin{figure}
\label{w_ws_pol}
\centerline{
\includegraphics[width=6.0cm,angle=-90]{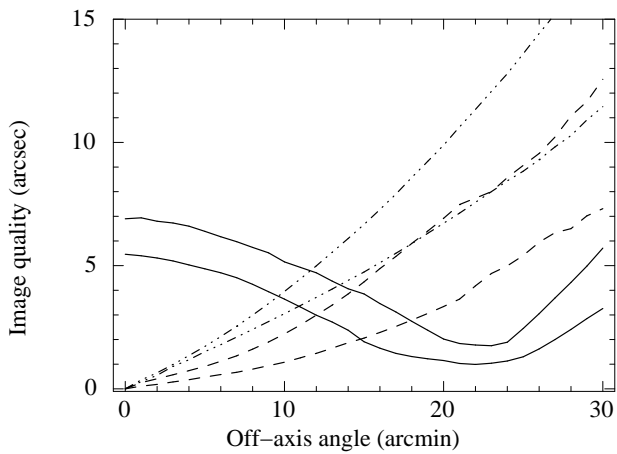}}
 \caption{Image quality functions over the field of view for different mirror shells with $f$=5 and $l=7$.
Dot-dot-dot-dashed lines refer to the Wolter I, dashed lines to Wolter-Schwarzschild and continuous lines
to the polynomial mirror. The two different lines for each mirror type are (from bottom to top),
HEW and $80\%$ EEF.}
\end{figure}

\section{Optimization of a mirror module}

As for the single shell optimization we need first to specify how the mirror shells are packed together, i.e. 
the acceptance angle (see also above). We decide to take the same acceptance angle for all the mirror shells,
simplifying the problem. To select this we compute the mean effective area of a mirror module for a few different 
idealized telescopes. In more details, a mean effective area over the entire field of view of $30'$ radius is 
computed as a function of the acceptance angle for a few idealized telescopes at two different energies 
(1 and 4 keV) shown in Fig. 5.
It is apparent from Fig. 5, where a few of these cases are shown, 
that the peak of the effective area is achieved always around $17'$.  
This angle basically depends on the field of view to be optimized, being $\sim 60\%$ of its radius.

\begin{figure*}
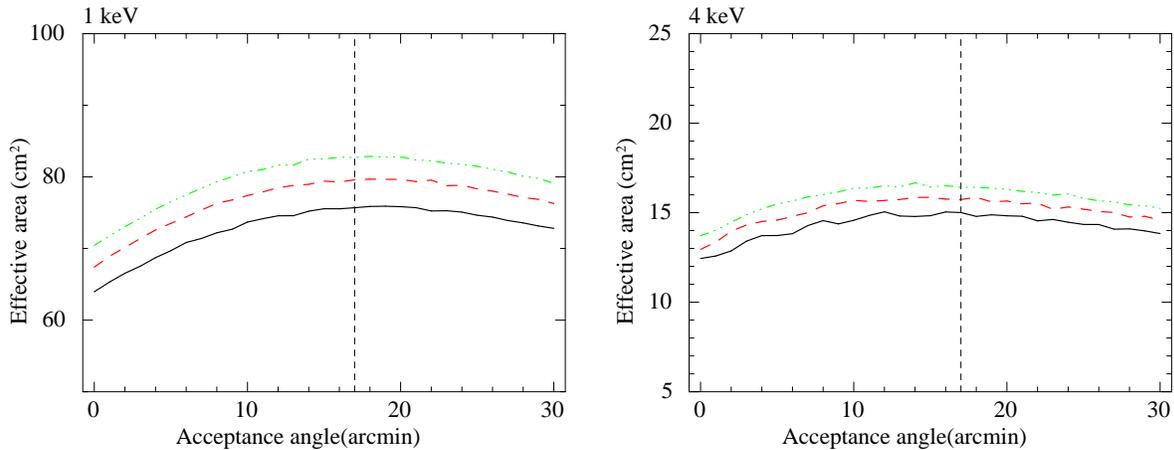

\label{accept}
\centerline{
\begin{tabular}{cc}
\includegraphics[width=6cm,angle=-90]{acc1.ps}&
\includegraphics[width=6cm,angle=-90]{acc2.ps}\\
\end{tabular}}
\caption{Effective areas for different telescope designs as a function of the acceptance angle. 
Left panel refers to areas at 1 keV and the right panel to 4 keV.
Dashed line is for a constant length mirror shells with $f=5$ and $l=8.5$, continuous lines for a 
$f=5$ and $l=8.5$ for the outermost shell and with mirror lengths scaling as $-0.55$, dot-dot-dot-dashed 
lines refer to $f=5$ and $l=6$ for the outermost shell and with mirror lengths scaling as $-0.2$. 
In all cases the best acceptance angle is around 17 arcmin off-axis.}
\end{figure*}

To proceed we have to specify a scaling factor for the thickness of the mirror shells.
Given for the outermost shell $D_0/T_0=650$ (with $T_0$ the thickness of the outermost shell) the thickness of any shell can 
be derived from $T={\sqrt{D\,D_0}\over{650}}$ (assuming therefore a scaling law with exponent of $0.5$, conservative with 
respect to the usual linear scaling).
This prescription does not affect the angular resolution of the mirror module but only its effective area 
(simply because the spacing among different shells is changing) and total weight.

Mirrors shells with the same focal length have focal planes
with different curvatures (Eq. 3). Therefore, if we align the mirror shells on-axis,
for any off-axis position the best focal plane will result from the average
of each shell focal plane. This results in an image blur
over the entire field of view. As noted in CC, one has to
build mirror shells which have the same curvature. As can be seen from Eq. \ref{sagitta}, decreasing the 
mirror diameter $D$ one has to decrease the mirror length $L$, scaling as $L\propto D^{1.8}$ (keeping the 
same focal length). This gives to the total mirror assembly a ``butterfly''-like shape.

In addition, images produced by different mirror shells do not superpose exactly, having different plate scales 
they focalize the relative spots with a focal offset relative to one another that increases with the off-axis angle. 
This problem can be overcome by assembling the mirror shells at different intersection planes, i.e. shells
must be moved relative to each other (see CC).

For the optimization of a mirror module we are not only interested in the angular resolution provided
by the mirror assembly but also to have the best resolution where there is also a larger effective area.
This implies that we have to weight the merit function (see Eq. 1) with the dependence of the effective area
on the off-axis angle $A_{\rm eff}(E,\,\Theta)$ (i.e. the vignetting function).
In addition, we evaluate this mirror effective area at two different energies (1 and 4 keV) in order to 
balance the low and high energy response. Because wide field X--ray mirrors have 
large incidence angles their effective area is limited in energy up to 4--6 keV. To increase the effective area
at higher energies we selected as best coating platinum (Pt). Platinum can be easily deposited onto the mirror carrier
through magnetron sputtering without degrading the surface microroughness. A thin layer of carbon (C) deposited on 
top of platinum (as well iridium or gold reflecting surfaces) improves the response at low energies (Pareschi et al. 2004).
As shown by Eq. \ref{hewsingle} the optimization procedure by itself promotes very low values of $f$ and $l$, 
but by including the manufacturing errors this is not the case anymore. For this reason we sum in quadrature 
to the image quality term an error term depending on energy ($\sigma(E)$). This term
includes the contributions from the surface roughness scattering, the mirror shell integration, the mirror length 
(shorter shells are more difficult to be built, etc.). 
Clearly a root-sum-squaring of a constant term is a simplification of the reality but as long as this is small it 
should not interfere with the minimization process.
We cover the $30'$ radius field of view with 9 CCDs (20 by 20 arcmin each) displaced in an inverse cut pyramidal shape, 
the tilting of which is part of the optimization procedure, to follow the curvature of the best focal plane.
One additional practical problem that one has to face is that adopting the scaling relation of Eq. \ref{sagitta} 
(in order to have the same curvature for all the mirror shells) one gets $D\propto L^{1.8}$. 
A further caveat concerns the manufacturing of mirror shells. As noted above too short mirror shells cannot be built 
without introducing large shaping errors which will basically nullify the polynomial design. 
We set a limit on the total shell length of 200 mm.  

Two different solutions are possible: 1) limit the number of shells and strictly follow the $D\propto L^{1.8}$ prescription
or 2) change somewhat the scaling of $D$ with $L$, introduce some aberrations, but obtain a larger effective area, especially
at high energies (e.g. $>4$ keV). We explore these two possibilities below.

\subsection{Best angular resolution}

For this option we simply scale the optimized single mirror shell according to
the $D\propto L^{1.8}$ scaling, keeping the same focal length for all mirror shells. 
The steep proportionality index between $D$ and $L$ is such that 
a (relatively) small number of shells can be accommodated. To show our results we choose to fix the
focal length $F=1000$ mm and scale the rest of the telescope according to $f$ and $l$. Taking $F=10000$ mm 
and scaling the other quantities accordingly, one obtains exactly the same angular resolution over the field of view
and an area larger by the square of the focal ratios (i.e. 100 times larger). Clearly the linear dimensions 
will be 10 times larger and the total weight 1000 times larger.
Within this 1000 mm focal length telescope we can accommodate 39 mirror shells with the outer 
(inner) shell of 200 (125) cm diameter. We plot in Figs. 6 and 7
the overall performances of such a telescope.
The weighted mean HEW is 2.3 arcsec (the root mean square, rms, spot is 1.7 arcsec and the 
$80\%$ EEF is 4.1 arcsec, see Fig. 6);
the effective area of one mirror module is shown in Fig. 7.

\begin{figure}
\label{sr_res}
\centerline{
\includegraphics[width=6.0cm,angle=-90]{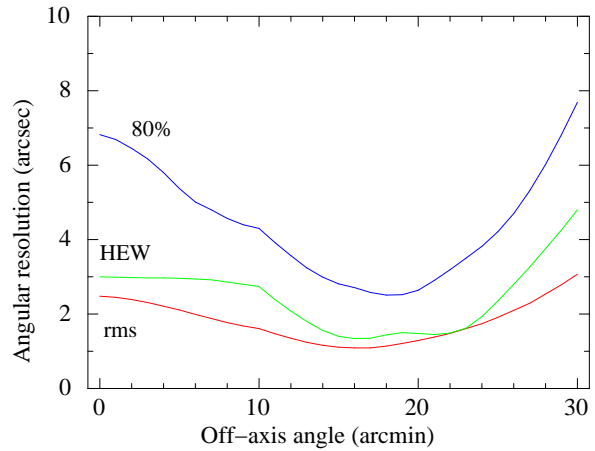}}
 \caption{Image quality functions evaluated at 1 keV over 9 CCDs in an inverse pyramidal shape 
as a function of the off-axis angle for a ``best angular resolution'' ($D\propto L^{1.8}$) telescope. 
The outermost shell values are $f=5$ and $l=7$. The shell length scales as above and the $f$ number 
(that has to change in order to have the same curved focal plane for all the shells) for the innermost shell 
is $f=8$. The edges seen in the curves are due to the change from one CCD to the next one (with different inclination).
These curves were obtained without considering scattering due to microroughness.}
\end{figure}

\begin{figure}
\label{sr_area}
\centerline{
\includegraphics[width=6.0cm,angle=-90]{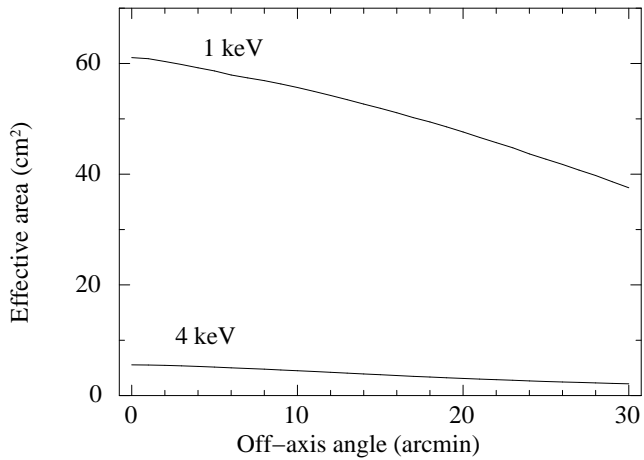}}
 \caption{Effective area of the unitary telescope described in the text (a focal length $F=1000$ mm)
at 1 and 4 keV for the ``best angular resolution'' ($D\propto L^{1.8}$) configuration.}
\end{figure}

\subsection{Compromise between best angular resolution and effective area at high energies}

As shown above, the request of scaling the mirror length as $D\propto L^{1.8}$ severely constraints 
the number of usable shells (below a given length it becomes increasingly difficult to manufacture 
mirror shells). A compromise would be to change the mirror length and the focal length 
of each shell according to Eq. \ref{sagitta}. 
Given the assumptions above we found a very simple and useful approximated formula,
analogue to  Eq. \ref{hewsingle}, to compute the expected mean HEW (weighted for the field of view and the effective area) 
for a mirror assembly with $f$ in the interval 5--17. Each shell will contribute to the total HEW with a weight that 
scales as $\sim f^{-2}$, resulting in:
\begin{equation}
{\rm HEW}=\sum_{k=5}^{17} A/k \times l_{f=5} \times (5/k)^{B} - C
\label{hewmirror}
\end{equation} 
\noindent where $A=0.478\times RF(f_k,\,E)/(RF(f_k,\, 1 \,{\rm keV})$ is the ratio 
of the reflectivities $RF$ of the mirror surfaces at a given energy $E$ to the reflectivity at 1 keV for a given focal ratio 
$f_k$ of the $k$ shell (so that if we are working with a monochromatic energy of $E=1$ keV  $A(k)= const = 0.478$); 
$B$ is the exponent scaling the length of the mirror shells ($l/l_0=(D/D_0)^{-B}$, with $l_0$ and $D_0$ the values of the 
outermost shell, if $B=0$ the mirror length is constant), $C=0.3$ and $l_{f=5}$ is the value of $l$ for the mirror shell
with $f=5$.  A similar expression holds at 4 keV with $A=0.507$.
This approximated formula predicts the global HEW of a mirror module with an accuracy better 
than $10\%$. One can use this empirical formula in an inverse way to optimize the mirror module.
Fixing the mean HEW to, e.g., 3.5 arcsec at 1 keV and an initial value for $f$ (we take $f=5$) one can derive the interval 
of mirror lengths again starting from a given value (see Table \ref{scaling}).

We then perform the optimization of the telescope. We need to start with the outermost shell for which we choose $f=5$ and 
several values of $l$. We start selecting an initial value for $l$. If we suppose an error budget of $\sim 2$ arcsec for any single shell 
from Eq. 2 we have that a mirror shell with $f=5$ should have $l\gsim 6$, because shorter shells will provide values 
much smaller than the error considered.
We then consider as a reference a mean HEW at 1 keV of 3.5 arcsec over the field of view. With this value we derive the 
scaling index $B$ from Eq. \ref{hewmirror}. For $l>9$ we end up with too large values of $B$ which will result in too short 
inner mirror shells. Thus we also exclude this interval, remaining with $l=6$, $l=7$, and $l=8$ for the outermost shell.
These three different telescopes are completely defined by the outer shell, by the scaling laws on $l$ and mirror 
thickness\footnote{We adopted a scaling of --0.5. In the past a linear decrease or a constant scaling has been adopted. 
As a check we also optimize along the lines described above  
a mirror module with a constant scaling law, obtaining very similar results. 
In particular, whereas the resolution is practically the same,
the area changes somewhat because of the different spacing between the single mirror shells.
The effective areas at 1keV and 4 keV at different off-axis angles are always within $5\%$
for the two configurations.
We also verified that, parametrizing the contribution of the $D/T$ in the error budget term $\sigma$ of Eq. \ref{meritmirror}
with a power-law, a scaling relation with index in the range 0.5--0.6 provides 
the best results, even if the dependence on the scaling index is relatively weak.},  
by the acceptance angle and by the shifts to be applied at the intersection plane to compensate for the different focal lengths 
(see also below). These telescopes can accommodate 78 mirror shells with the outer (inner) shell of 200 
(61) cm diameter. These telescopes are then optimized with a DLS with a merit function similar to the one used for the single
mirror shell. This provides the starting point for the hammer optimization. We ray-trace the DLS optimized telescopes and 
find out the spot barycentric positions at 1 keV for an interval of off-axis angles (namely, from $0'$ to $30'$ in step of $5'$). 
We then fix these positions and optimize shell by shell each telescope with the hammer optimization.
In Table \ref{scaling} we provide the mean weighted values for these three configurations, which are very similar, and 
in Figs. 8 and 9 
we plot the HEW and the effective area at 1 keV and 4 keV for the three configurations, respectively. 

To finally select the best telescope we envisage a merit function including also the effective area of the form
\begin{eqnarray}
&&M_{\rm mm} = \sum_{i=0}^{n} \, \Bigl(  \nonumber \\
&& \sum_{j=0}^{m}\, \sqrt{(HEW(E_i,\, \Theta_j)+EEF_{80}(E_i,\, \Theta_j))/2)^2+\sigma(E_i)^2} \times \nonumber \\
&& \times A_{\rm eff}(E_i, \,\Theta_j) \times \Theta_j d\Theta_j  \Bigr) /  \nonumber \\
&& /  \sum_{j=0}^{m}\, A_{\rm eff}(E_i, \,\Theta_j) \times \Theta_j d\Theta_j 
\label{meritmirror}
\end{eqnarray}
  
All these configurations provide a mean HEW over the field of view of $\sim 3$ arcsec at 1 keV and $\sim 4$ arcsec at 4 keV,
the underlying hypothesis being that the intrinsic total error $\sigma$ on the mirror fabrication and integration 
is relatively small, at a level of $3-3.5$ arcsec (providing an effective HEW $\sim 5$ arcsec, summing in quadrature the HEW 
and the error). No surface microroughness has been considered.
For larger errors, $\sigma\gsim 5$ arcsec, all the benefits of the polynomial design are smeared out by the 
error (and the optimization processes described here will have to be revised). 
The merit function $M_{\rm mm}$ selects the $l=8$ as the best telescope, even if differences among 
the three telescopes considered are small (see Figs. 8 and 9).

\begin{table*}
\caption{Summary of the three telescopes (with initial values $l=8$, $l=7$ and $l=6$ discussed in the text) image quality 
characteristics$^+$ (obtained by assuming implicitly an overall mirror error at a level of 3--3.5 arcsec).}
\label{scaling}
\begin{tabular}{cccccccc}
\hline
            &           &               &          & 1 keV  &        &         \\
Initial $l$ & $B$ index & $l$ range     & $f$ range&Mean rms&Mean HEW&Mean EEF $80\%$ & Mean area\\     
            &           &               &          &(arcsec)&(arcsec)& (arcsec) & cm$^2$\\
\hline
        8   &  0.565    &  8.00 -- 4.30 & 5--17    & 2.69 & 3.35   & 6.19 & 74.2\\
        7   &  0.335    &  7.00 -- 4.84 & 5--17    & 2.75 & 3.04    & 5.99 & 73.0\\
        6   &  0.085    &  6.00 -- 5.46 & 5--17    & 2.84 & 2.97    & 5.93 & 71.6\\
\hline
            &           &               &          & 4 keV  &        &       &  \\
\hline
        8   &  0.565    &  8.00 -- 4.30 & 5--16    & 3.68   & 4.01   & 8.12 & 14.9\\
        7   &  0.335    &  7.00 -- 4.84 & 5--16    & 4.13   & 4.13   & 8.66 & 14.9\\
        6   &  0.085    &  6.00 -- 5.46 & 5--16    & 4.36   & 4.47   & 9.26 & 14.9\\
\hline
\end{tabular}

\noindent $^+$ obtained by assuming implicitly an overall mirror error at a level of 3--3.5 arcsec.
\end{table*}

\section{Building a wide-field X--ray telescope}

We describe here the optimization of a wide field X--ray telescope. Given the large freedom we have to set a number of 
boundary conditions for this instrument. The rationale is to have an instrument able to carry out deep survey of the 
X--ray sky therefore, in addition to a good image quality over large field of view (30 arcmin radius), we need also a large effective area.
To set the stage we fix for the effective area at 1 keV a lower limit of 8,000 cm$^2$ (not-including detector 
efficiency and filter transmission) and 2,000 cm$^2$ at 4 keV. We also require a mean HEW over the field of view lower than 5 
arcsec at 1 keV. The total diameter of the mirror assembly should be smaller than 2500 mm diameter and the focal length 
shorter than 6000 mm to fit in within an ATLAS V402 shroud. The total weight of the mirrors should be less than 1200 kg. 

In the fabrication of a real telescope we have to account for a number of errors ranging from the mirror fabrication, 
integration, alignment as well as scattering from the mirror surface roughness (especially important at high energies). 
The experience with the Chandra satellite shows that it is possible to manufacture mirrors having angular resolution 
$\lsim 1$ arcsec as well as a very low surface roughness ($\sim 3$ \AA\ r.m.s. Weisskopf 2001). 
These values were obtained with very thick mirrors ($\sim 20$ mm, resulting in $D/T\sim 50$) and ratio of mirror 
length over mirror diameter $2\,L/D$ $>1$. In addition a fiducial light system is able to track the satellite attitude 
better than $0.5''$. 
The technological challenge is therefore to build thinner and shorter mirror shells able to meet the requirement described 
above. Given the design of one single module discussed before, we have about 4 arcsec error 
for the manufacturing of the mirror shells (to be summed in quadrature to the intrinsic HEW of the optical design). 

\begin{figure}
\label{hew_l876}
\centerline{
\includegraphics[width=6.0cm,angle=-90]{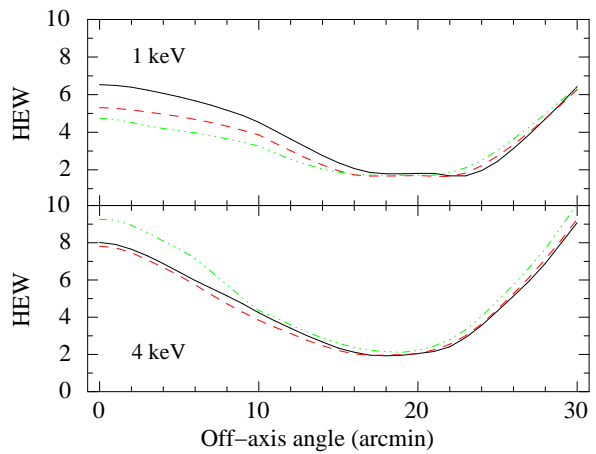}}
\caption{HEW for the three designs discussed as a function of the off-axis angle. In the top panel are shown 
the HEW at 1 keV: the continuous line refers to the $l=8$ (initial value for the outermost shell) design, the dashed 
line to $l=7$, and dash-dot-dotted line to $l=6$. In the lower panel the same curves are shown for an energy of 4 keV.
All the curves were obtained without considering scattering due to microroughness.}
\end{figure}

\begin{figure}
\label{area_l876}
\centerline{
\includegraphics[width=6.0cm,angle=-90]{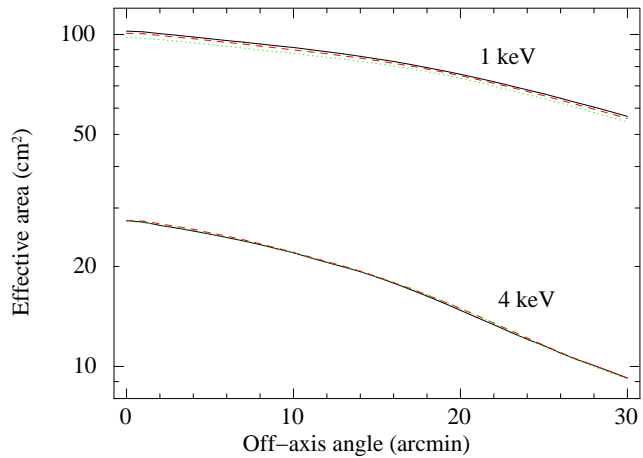}}
\caption{Effective areas of the unitary telescope (focal length $F=1000$ mm) for the three designs discussed in the text at 
two different energies over the field of view at 1 keV and 4 keV. Continuous lines refer to $l=8$ (initial value for 
the outermost shell) design, dashed lines to $l=7$ and dotted lines to $l=6$. }
\end{figure}

Our best design consists of three telescopes with focal length of 5500 mm and diameter for the outermost mirror 
shell of 110 cm. This corresponds to a scaled version of $l=8$ telescope.
The telescope characteristics are summarized in Table \ref{telescope}. 
The angular resolution of the telescope is as in Figs. 8 and 10.
The mirror effective area (including a $10\%$ 
obstruction due to the support structure) of the three modules is shown in Fig. 11.

\begin{table}
\caption{Characteristics of the proposed surveying telescope.}
\label{telescope}
\centerline{
\begin{tabular}{cc}
\hline
Focal Length (cm) & 5500 \\
Number of Optics Modules & 3\\
Number of shells & 82 \\
Radius [MAX - min]  & 55.0 -- 16.5 (cm)\\
Total Length [MAX - min]  &  44.0 -- 23.5 (cm)\\
Thickness [MAX - min]  & 2.2 -- 1.2 (mm)\\
Total on-axis Effective Area$^*$ (1 keV) & 9236 (cm$^2$)\\
Total on-axis Effective Area$^*$ (4 keV) & 2565 (cm$^2$)\\
Total Weight (3 modules) & 930 (kg) \\
\hline
\end{tabular}
}

$^*$ this area refers to the total mirror area for the three modules
and accounting for a $10\%$ obstruction from the support structure.
 
\end{table}

Given that we specified a thickness for the mirror shells, we also need to prove that these numbers are realistic. 
Therefore we will now explore what we think is the most appropriate approach to build such a telescope.

The realization of mirror shells with a small aspect ratio ($l$ ratio $\sim 3$ times lower than XMM-Newton and Chandra) 
implies an increased difficulty in reaching very good angular resolution. This is due to various aspects, particularly relevant 
if replication approaches are utilized, for instance using electroformed Nickel as a carrier (as done with BeppoSAX, 
XMM-Newton and Swift XRT). In this case the mechanical behavior of the mirror shell is closer to a `belt-like' configuration 
rather than a `tube-like'. In particular, the degradation of the mirror shell stiffness due to its cross section `out of 
roundness' has a much higher impact (e.g. `dead area' from not well polished and figured sections at 
the mandrel intersection planes, `trumpet' effects due to internal stresses in Nickel replicas at the top and bottom of 
the shell, can cause important image quality worsening).
The angular resolution is strongly affected by the slope errors caused by out-of-phase azimuthal errors, linearly correlated 
to the mirror shell length.

\begin{figure}
\label{12spot}
\centerline{
\includegraphics[width=6.0cm,angle=0]{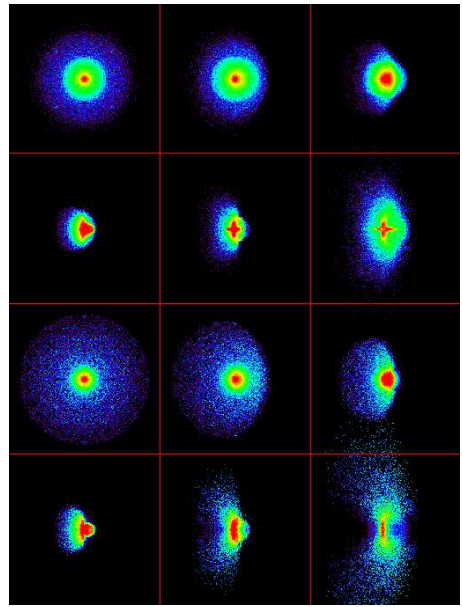}}
\caption{Full telescope spot at 1 keV (6 upper panels) and 4 keV (6 lower panels) at off-axis 
angles of $0'$, $6'$, $12'$, $18'$, $24'$ and $30'$, respectively, evaluated on the best 
(inverse pyramid) CCD focal plane. Surface roughness has not been considered. Images are 20 arcsec apart.
} 
\end{figure}

These problems can be solved assuming higher rigidity mirror shells, based on a material with very good 
mechanical properties such as Silicon Carbide (SiC) in combination with a direct figuring and polishing of the mirror 
shells.
SiC is a material well known for its outstanding thermal-mechanical properties for which it is ideal for the realization 
of high precision optics, in particular for space applications. It offers a number of advantages over other traditional 
optical substrate materials such as: low density (3.2 g cm$^{-3}$), low thermal expansion coefficient ($2.8\times10^{-6}$ 
K$^{-1}$), high thermal conductivity (3.7 W cm$^{-1}$ C$^{-1}$), high specific heat (0.7 J g$^{-1}$ C$^{-1}$), high modulus 
of elasticity (450 GPa). It has already been used for the realization of the optics for space based projects, like for 
instance the HERSCHEL 3.5 m primary mirror. 

The feasibility of large size wide-field grazing-incidence polynomial mirrors made in SiC has already been demonstrated in 
the past at our institute (Citterio et al. 1999; Ghigo et al. 1999), with the realization (produced via epoxy replication) of a 
few gold coated mirror shells. The best result has been achieved with a prototype having a 2 mm thickness, 60 cm diameter and 24 
cm height. 
It was based on a SiC Chemical Vapour Deposition (CVD) carrier, using a superpolished polynomial mandrel for replication.  
The mandrel profile was measured to have a HEW of 7 arcsec on average over the 1 degree diameter field of view. 
The shell had a HEW of $\sim 10$ arcsec, almost constant across a 60 arcmin wide off-set scan, as measured at Marshall Space Flight
Center (NASA) in X--rays (Citterio et al. 1999; Ghigo et al. 1999). 
It should be noted that Ni electroformed optics 1 mm thick (i.e. with the same weight as the one in SiC) produced from the same 
mandrel presented a much worse angular resolution (35 arcsec HEW across the field). 

\begin{figure}
\label{areatot}
\centerline{
\includegraphics[width=6.0cm,angle=-90]{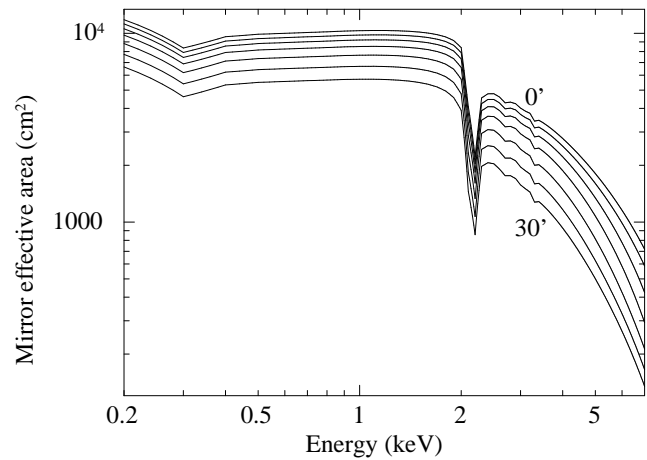}}
\caption{
Effective area for the full telescope (three mirror units) discussed in the text, including 
$10\%$ obstruction from the support structure. From top to bottom areas are for $0'$ to $30'$ 
off-axis angles in step of $5'$.} 
\end{figure}

In order to fulfill the stringent angular resolution requirement as well as maintain the weight of the 
mirror assembly to a reasonable value, mirror shells have to built in SiC. 
However, instead of using the replication approach, a deterministic direct polishing method should be 
used to manufacture the SiC mirror shells with the required optical quality.  
At this regard, the thickness trend of the mirror shells along the series of diameters has been chosen in such a way to maintain 
almost constant stiffness. With the present set of SiC wall thicknesses the deformations introduced by an astatic support system, 
to be used to hold the mirror shells during the polishing and metrology operations, are negligible (less than 0.5 arcsec HEW), 
as we have verified with a careful Finite Element Method (FEM) analysis.

More in details, the SiC mirror carriers can be produced using the CVD process, well developed and available from a number of US 
and European companies. The carriers should then be figured and polished making use of a `deterministic' method.
This implies that after the measurements of the actual profile of the mirror shell to be polished, a corrective matrix is 
determined and supplied to a computer numerical controlled polishing machine which provides the corrective action according 
to the given error matrix. In a few iterations it should be possible to reach the required specifications.
On the market there are available CNC polishing machines that through a combination of `jet-polishing' and `bonnet-polishing' processes
can provide the requested characteristics.
The `jet-polishing' approach is currently used in a number of high precision optics productions, including the realization 
and development of collectors for nanolithography and the segmented mirrors for large size ground based telescopes like European
Extremely Large Telescope (E-ELT). 

Once more, in order not to destroy the image quality obtained from our optical design, fabrication and integration errors should be 
at the level of $\lsim 4$ arcsec while the errors induced by the space thermal-mechanical environment must be less than 1 arcsec.

\section{Discussion and conclusions}

Images taken with the COBE and WMAP satellites of a Universe only a few hundred thousand years old
show a nearly uniform picture with only subtle features: the seeds of larger structures yet to
come. Hubble Space Telescope and 10-meter class telescopes peering deep into the Universe
to within a few billion years after its birth show already fully-formed galaxies. Local galaxy
surveys show that most galaxies are clustered in physically bound units (clusters and groups)
that on large scales map a web-like distribution of matter. Still, a number of open questions
remain on the structure of the Universe at high redshifts, its formation and evolution, its relation
with the growth of supermassive black holes that power Active Galactic Nuclei (AGN). 

Optical surveys are extremely limited in their ability to answer our questions on the distant
Universe and high-energy phenomena. Consider, for example, the Sloan Digital Sky Survey (SDSS),
which will contain hundreds of millions of galaxies, most at redshifts less than 0.5. Even
in this massive survey, distant objects such as clusters of galaxies at redshifts near 1 are
confused in the general background and foreground of faint galaxies. Therefore they are difficult to locate
through the `fog' of the foreground population. On the contrary, a unique feature of the X--ray
sky is its low background and the preponderance of faint objects at substantial cosmic distances.
This makes it possible to accurately select groups and  clusters of galaxies and AGN directly from
the X--ray images, and use them to trace the matter distribution on cosmological scales. X--ray
groups can be used to map structures at low redshift.

The ability of X--ray images to peer deep into the early Universe is also a key in providing a way to study the
evolution of classes of objects over cosmological time-scales (for a broad perspective see the WFXT white papers submitted to
the 2010 Decadal Survey, Murray et al. 2009; Giacconi et al. 2009;  Vikhlinin et al. 2009; Ptak et al. 2009). In a large 
survey, these issues can all be addressed in a statistically significant way, in order to precisely constrain theories. 
At the same time, extremely powerful and rare phenomena within our own Galaxy, become
evident only when large volumes are explored in the high-energy domain of X--rays.
A deep (nearly) all-sky X--ray survey with high spatial resolution is therefore a natural, necessary step to complement
and significantly extend the optical sky surveys as the classic Palomar or SDSS sky atlas as well as radio and near-infrared
already available or available in the next decade.

In this paper we focus on the optimization of the optical design in relation to the scientific drivers of 
the WFXT mission. In particular, we describe step by step 
the design of a wide field X--ray telescope tailored for surveying 
the X--ray sky. This is a clear step forward with respect to previous papers (see BBG and CC).
A general indication of the manufacturing process of the mirrors have been presented, while a detailed 
description of the latter aspects will be presented in a forthcoming paper.

\section*{Acknowledgments}
We warmly thank R. Giacconi, S. Murray, R. Elsner, and M. Weisskopf for useful discussions.
We thank P. Span\`o for useful discussions on the Zernike polinomials and {the referee for useful 
comments}.

\appendix
\section{Aberration analysis}

\begin{table*}
\caption{Results of the aberrational analysis for three single mirror shell of different telescope types: Wolter I (Wol. I),
Wolter-Schwarzschild (WS) and Polynomial (Pol). For the different telescopes are reported the spot rms in arcsec.}
\label{opd}
\begin{tabular}{cccccccc}
\hline
Aberration$^*$ &Zernike pol. order$^+$ & Wol. I ($0'$) & WS ($0'$) & Pol. ($0'$) & Wol. I ($20'$) & WS ($20'$) & Pol. ($20'$)\\
\hline
Initial         & -- & 0.0           & 0.0       & 3.14        & 4.56           & 3.50       & 2.37        \\
Piston and tilt &(0,0) (1,1) (1,--1)& 0.0           & 0.0       & 3.14        & 4.56           & 3.50       & 2.37        \\
Astigmatism    &(2,2) (2,--2)& 0.0           & 0.0       & 3.14        & 3.81           & 2.42       & 0.93        \\
Coma           &(3,1) (3,--1)& 0.0           & 0.0       & 3.14        & 2.34           & 2.42       & 0.93        \\
Trefoil        &(3,3) (3,--3)& 0.0           & 0.0       & 3.14        & 2.34           & 2.42       & 0.93        \\    
3$^{rd}$ spheric.&(4,0)      & 0.0           & 0.0       & 1.09        & 0.36           & 0.92       & 0.93        \\
Second. astigm.&(4,2) (4,--2)& 0.0           & 0.0       & 1.09        & 0.08           & 0.81       & 0.87        \\
Quadrifoil     &(4,4) (4,--4)& 0.0           & 0.0       & 1.09        & 0.08           & 0.81       & 0.87        \\    
5$^{rd}$ spheric.&(6,0)      & 0.0           & 0.0       & 0.43        & 0.07           & 0.81       & 0.74        \\
\hline
\end{tabular}

\noindent $^*$ The values refer to the rms spot (in arcsec) after removal of the corresponding aberration.

\noindent $^+$ Zernike polynomials are characterized by a primary number $n$ and a secondary number $m$
which can vary only among $-n\, n$ integers and take only odd (even) numbers if $n$ is odd (even). 

\end{table*}

The idea underlying the polynomial design is that off-axis performances can be improved by introducing
(mainly) spherical aberration on-axis. 
To deepen our understanding of the aberrations we carry out a novel analysis based on (suitable) Zernike 
polynomials. Zernike polynomials are a sequence of polynomials that are orthogonal on the unit disk and 
are widely used in optics. They are widely used to study optical systems' aberrations expanding the 
aberration function in series. These polynomials need some modifications to be applied to X--ray astronomy, where the 
exit pupil is not an unitary disk but (in this case) a thin annular region in the range 
0.989 to 1 (being 1 the length of the outer radius).
    
We adopted similar polynomia defined however on an unitary annulus instead of an unitary disk  (Mahajan 1981). 
We first analyze the on-axis image of a Wolter 
I mirror shell to be sure that our analysis does not introduce spurious effects. As expected the Zernike 
analysis shows a spot without aberrations. We then consider an off-axis angle of $20$ arcmin. As expected 
the Wolter I image is severely affected by aberrations. The rms spot obtained by a ray-tracing simulation is 
3.89 arcsec. We evaluate with the Zernike annular polynomials the expected optical path-length difference (OPD) 
rms spot.
We then decompose the OPD into the contribution of the different aberrations (see Table \ref{opd}).
We repeat the same analysis for a Wolter-Schwarzschild mirror shell and for our optimized polynomial mirror shell
in the $f=5$ and $l=7$ configuration (see also Fig. 4).
In the case of the polynomial
mirror shell at $20'$ off-axis we see the presence of large contribution from astigmatism 
(nearly $80\%$ if summed in quadrature) some $10\%$ contribution from higher orders aberrations. 
This is at variance with the Wolter I OPD where the main contributions come from spherical aberration and come. 
In the case of the Wolter-Schwarzschild shell at large off-axis angles we see the presence of comparable 
contributions from spherical aberration and astigmatism.

\section{Scaling and shifting mirror shells}

The need for a change in the mirror shell focal lengths and for moving their intersection planes with respect to 
each other can be explained as following. The plate scale of a telescope relates the angular scale in the sky  
to the linear scale on the detector (and can be measured in mm/arcsec). For an optical telescope the plate scale 
$PS$ is simply $\arctan{1/F}$. For a grazing incidence X--ray telescope where the first focal surface forms an angle 
$\alpha$ with the optical axis the plate scale on-axis is:
\begin{equation}
PS=\arctan(\cos^2(4\,\alpha)/F) \label{pson}
\end{equation} 

\begin{figure}
\label{spot}
\centerline{
\includegraphics[width=6.0cm,angle=-90]{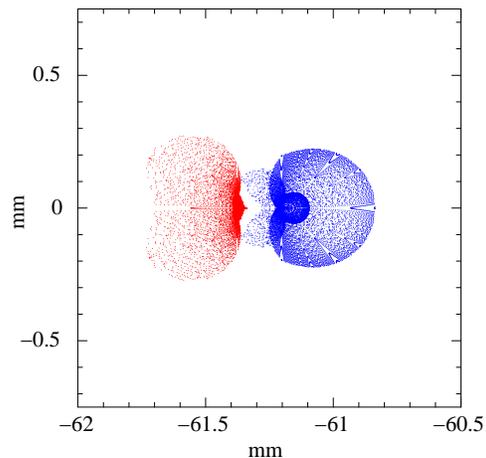}}
\caption{Image spot of two different shells (outermost and innermost) at an off-axis position of $20'$
at 1 keV taking the same focal length for the two shells. 
This blur is corrected changing the focal length of the innermost shell and then shifting it at the intersection plane
to achieve the same plate scale as the outermost.
the shift at the intersection plane. In this configuration 1 mm in the focal plane corresponds to 20 arcsec.}
\end{figure}

\begin{figure}
\label{moving}
\centerline{
\includegraphics[width=6.0cm,angle=-90]{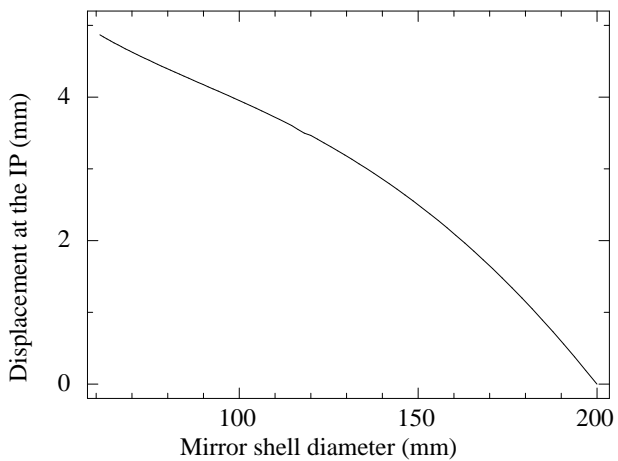}}
\caption{Displacements at the intersection plane (IP) in mm to align each mirror shell spot on the 
inverse pyramid focal plane as a function of the mirror entrance diameter. Numbers refer to the
unitary telescope with $F=1000$ mm. 
} 
\end{figure}

It is apparent from Eq. \ref{pson} where the spot produced by the innermost and outermost shells 
are ray-traced at an off-axis position of $20'$ are shown, that 
considering different mirror shells we have different incidence angles $\alpha$ and therefore different $PS$,
leading to an image blur. This can be compensated by changing the focal length $F$. However each mirror shell will then 
focus the image on a different focal plane, which can be compensated by a shift of the shell at the intersection plane.

In order to underline once more the importance of the additional corrections we introduced, we 
show in Fig. B1 the two spots focalized by the innermost and outermost mirror shells of the telescope 
described above. They have the same focal length and are not shifted one with respect to each other. 
Clearly not correcting for this effect produces an increase in the spot size at any energy and a shift in the 
barycenter of the image at different energies, since innermost mirror shells contribute more at higher energies
with respect to the outermost shells that contribute more at the lower energies. 
As already noted in CC (and BBG) this problem can be overcome by making mirror shells with different intersection 
planes, i.e. shells must be translated relative to each other. In the $F=1000$ mm focal length telescope, mirror shells 
have to be shifted by up to 5 mm at the intersection plane. The scaling is not linear but we obtain a good fit with a 
cubic function (see Fig. B2). 

\end{document}